\documentclass[11pt]{article}

\usepackage{color,graphicx}
\usepackage{young}
\usepackage[vcentermath]{youngtab}
\usepackage{amsmath,amssymb,graphicx}
\usepackage{hyperref}
\definecolor{darkred}{rgb}{0.65,0.15,0}
\hypersetup{pdfborder={0 0 0},colorlinks=true,urlcolor=darkred,citecolor=blue,linkcolor=darkred,linktocpage=true}

\usepackage{cite}
\usepackage{amsmath}
\usepackage{amsfonts}
\usepackage{amssymb}
\usepackage{graphicx}%
\usepackage{amsthm}
\usepackage{mathrsfs}
\usepackage{lscape}
\usepackage{multirow}
\usepackage[T1]{fontenc}
\usepackage{enumerate}
\usepackage{color}
\usepackage{hyperref}
\hypersetup{
   colorlinks   =  true,
    citecolor    = red,
     urlcolor	=magenta,
}

\def\4diml{four-dimensional}

\def\-1{^{-1}}

\newcommand{\M}{\mathscr{M}}
\newcommand{\D}{\mathscr{D}}
\newcommand{\G}{\mathscr{G}}

\makeatletter

\@addtoreset{equation}{section}
\makeatother

\begin{document}

\thispagestyle{empty}

\vspace{5mm}

\begin{center}
{\LARGE \bf Yang-Baxter deformations of the $GL(2,\mathbb{R})$ WZW model \\[2mm]  and non-Abelian T-duality}

\vspace{14mm}
\normalsize
{\large   Ali Eghbali\footnote{eghbali978@gmail.com}, Tayebe Parvizi\footnote{t.parvizi@azaruniv.ac.ir},
Adel Rezaei-Aghdam\footnote{Corresponding author: rezaei-a@azaruniv.ac.ir}}

\vspace{4mm}
{\small {\em Department of Physics, Faculty of Basic Sciences,\\
Azarbaijan Shahid Madani University, 53714-161, Tabriz, Iran}}\\


\vspace{10mm}

\begin{tabular}{p{12cm}}
{\small

By calculating inequivalent classical r-matrices for the $gl(2,\mathbb{R})$ Lie algebra as  solutions of (modified) classical  Yang-Baxter equation
((m)CYBE)), we classify the YB deformations of Wess-Zumino-Witten (WZW) model on the $GL(2,\mathbb{R})$ Lie group in twelve inequivalent families.
Most importantly, it is shown that each of these models can be obtained from a
Poisson-Lie T-dual $\sigma$-model in the presence of the spectator fields when the dual Lie group is considered to be
Abelian, i.e. all deformed models have Poisson-Lie symmetry just as undeformed
WZW model on the $GL(2,\mathbb{R})$. In this way, all deformed models are specified via spectator-dependent background matrices.
For one case, the dual background is clearly found.

}
\end{tabular}
\vspace{6mm}
\end{center}

\newpage
\setcounter{page}{1}

\tableofcontents

 \vspace{5mm}
 \vspace{5mm}

\section{\label{Sec.II} Introduction}
The deformation of integrable two-dimensional $\sigma$-models has attracted considerable attention in two decades ago, in particular given
their applications in string theory and AdS/CFT \cite{klimcik2003yang,klimvcik2009integrability,klimvcik2014}
(for a comprehensive review, see \cite{Hoare}).
Integrable deformations of $SU(2)$ principal chiral model were firstly presented in
\cite{family,Relativistically,representation}.
The generalization of \cite{Relativistically,representation} as YB (or $\eta $) deformation of chiral model
was introduced by Klimcik in \cite{klimcik2003yang,klimvcik2009integrability,klimvcik2014}.
The YB deformations are based on $\mathbf{R}$-operators satisfying
the (m)CYBE or CYBE (homogeneous YB deformations)
 \cite{matsumoto2015yang}. The application of these integrable deformations to $AdS_5 \times S^5$ superstring action has been presented in \cite{delduc2014integrable,Jordanian,supercoset}.
Note that the initial input for construction of a YB deformed background is classical r-matrix.
The $r$-matrices may be divided into Abelian and non-Abelian. It
has been proved that the YB deformed chiral models related to Abelian $r$-matrices correspond to T-duality shift T-duality transformations \cite{Tongeren}.
In the case of non-Abelian $r$-matrices it has been shown that the YB deformed chiral model corresponds to deformed  T-dual models (with invertible two cocycle $\omega$ such that $\omega^{-1}=\mathbf{R}$) \cite{Borsato}.
Some of the YB deformations of the WZW models with compact or noncompact Lie groups have been also performed in \cite{delduc2015integrable,kyono2016yang,Klimcik2017,Quantum,2020,Heisenberg}.
Generalization of this type of the deformations to Lie supergroups has been recently explored in \cite{EPR2}.

The main purpose of this paper is to construct the YB deformations of WZW model based on the $GL(2,\mathbb{R}) $ Lie group.
We first classify the inequivalent classical  $r$-matrices as solution of
(m)CYBE by using the automorphism transformation associated to the $gl(2,\mathbb{R}) $ Lie algebra.
Then we obtain the YB deformed backgrounds of the  $GL(2,\mathbb{R}) $ WZW model. As previously shown in \cite{Exact}, the WZW model on the $ GL(2,\mathbb{R})$
has Poisson-Lie symmetry with spectators. Here we will show that all YB deformed backgrounds have Poisson-Lie symmetry, in such a way that the resulting deformed backgrounds can be represented as original models of
Poisson-Lie T-dual $\sigma$-models in the presence of the spectator fields when the dual Lie group is considered to be
Abelian; in fact all deformed models will have Poisson-Lie symmetry just as undeformed WZW model on the $GL(2,\mathbb{R})$ \cite{Exact}.

The paper is organized as follows. In Sec. 2, we start by recalling the YB deformation of WZW model.
In Sec. 3, we first review the construction of WZW model on the $ GL(2,\mathbb{R})$ Lie group \cite{Exact},
then we solve the (m)CYBE in order to obtain the inequivalent classical r-matrices for the $gl(2,\mathbb{R})$ Lie algebra.
The backgrounds of YB deformed WZW models on the $ GL(2,\mathbb{R})$ Lie group are also constructed in this section;
the results are summarized in Table 1. The conformal invariance
conditions of the YB deformed backgrounds up to the one-loop order are discussed at the end of Sec. 3.
In Sec. 4, we show that the YB deformed models can be considered as original ones
of non-Abelian T-dual $\sigma$-models. For all deformed backgrounds, the spectator-dependent background matrices
are represented in Table 2.
At the end of Sec. 4, we also obtain the non-Abelian target space dual
for one case of the deformed models. Some concluding remarks are given in the last section.


\section{\label{Sec.II} A review of the YB deformations of WZW model}

The YB deformation of WZW model on a Lie group $G$ is giving by  \cite{{delduc2015integrable},{kyono2016yang}}
\begin{eqnarray}\label{2.5}			
S^{YB}_{_{WZW}}(g)=\frac{1}{2}\int_{_\Sigma}d\sigma^+ d\sigma^-  ~\Omega_{ab} J_{+}^{a} L_{-}^{b}  +\frac{\kappa}{12}\
\int_{_B} d^3 \sigma~
\varepsilon^{ \gamma \alpha \beta}~{\Omega}_{ad} \;f_{bc}^{~~d}~
L^{a}_{_\gamma} L^{b}_{_\alpha} L^{c}_{_\beta},
\end{eqnarray}
where $\Sigma$ is worldsheet with the coordinates $(\tau , \sigma)$\footnote{In the rest of the paper,
however, we will use the standard lightcone variables $\sigma^{\pm} = {(\tau\pm\sigma)}/\sqrt{2}$ together with $\partial_{\pm}=
(\partial_{\tau} \pm \partial_{\sigma}) /\sqrt{2}$.
Our conventions are such that the alternating tensor is $\varepsilon^{+-}=1$.}.
In the second integral, {\small $B$} is a three-manifold bounded by the worldsheet with the
coordinates $\alpha=(\xi, \tau, \sigma)$, and $\kappa$ is a constant parameter.
Here $\Omega_{ab}$  is a non-degenerate ad-invariant symmetric bilinear form on Lie algebra ${\G}$ of $G$ where is defined by $\Omega_{ab} =<T_a , T_b>$;
moreover, $f_{ab}^{~~c}$ stand for the structure constants of ${\G}$, and $ {L_{_{\pm}}} $ are the components of the left-invariant one-forms which are defined in the following way
\begin{eqnarray}\label{2.6}			
g^{-1} \partial_{_{\pm}}g=L_{\pm}=  L_{\pm}^{a}T_{a},
\end{eqnarray}
in which $g:\Sigma \rightarrow G$ is the element of Lie group $G$ and $T_{a}$ ($a=1,..,dim ~{G})$ are the bases of Lie algebra ${\G}$. Here the linear operator
{$\mathbf{R}: {\G} \rightarrow {\G}$  is a solution of (m)CYBE \cite{matsumoto2015yang}
	\begin{eqnarray}\label{2.3}
	[\mathbf{R}(X),\mathbf{R}(Y)]-\mathbf{R}\big([\mathbf{R}(X),Y]+[X,\mathbf{R}(Y)]\big)=\omega [X,Y],~~~~\forall X, Y \in {\G},
	\end{eqnarray}
where $\omega$ is a constant parameter. When $\omega=0 $, equation \eqref{2.3} is called the CYBE while for $\omega=\pm1$ this equation can be generalized to the mCYBE. The deformed currents $J_{\pm}$ are defined by means of the following relation \cite{delduc2015integrable}
	\begin{eqnarray}\label{2.8}
	J_{\pm} = (1+\omega \eta^{2})\frac{1 \pm \tilde{A} \mathbf{R}}{1-\eta^{2}\mathbf{R}^{2}} L_{\pm},
	\end{eqnarray}
where $\eta$ and $\tilde{A}$ are some real parameters measuring the deformation of WZW model.
If we set $\eta = \tilde{A}=0 $ and $\kappa=0 $ one then recovers the action of the
principal chiral model. Also for $\eta = \tilde{A}=0$ and $ \kappa=1$ the action reduces to the same  standard WZW model \cite{delduc2015integrable}.
Notice that the parameter $\omega $ can be sometimes normalized by rescaling $\mathbf{R}$, accordingly, one can consider  $\omega=0, \pm1 $.
 Indeed, the model \eqref{2.5} is integrable as shown in \cite{delduc2015integrable}. In the following we shall consider the
 model \eqref{2.5} for the  $ GL(2,\mathbb{R})$ Lie group.


\section{\label{Sec.III} YB deformations of WZW model on the $GL(2,\mathbb{R})$ Lie group}

Similar to the calculations performed to obtain the classical r-matrices of the $h_{4}$ Lie algebra \cite{Heisenberg},
here we use the automorphism transformation of the $gl(2,\mathbb{R})$ Lie
algebra to classify all corresponding inequivalent classical r-matrices as solutions of (m)CYBE.
In order to obtain the YB deformations of the $GL(2,\mathbb{R})$ WZW model, one
needs to calculate all linear $\mathbf{R}$-operators corresponding to the obtained classical r-matrices.
Then, by calculating the deformed currents $J_{\pm}$ we will obtain all YB deformations of the $GL(2,\mathbb{R})$ WZW model.
Before proceeding to obtain these, let us first consider the undeformed WZW model on the $GL(2,\mathbb{R})$.

\subsection{\label{subSec.III.1} The WZW model based on the $GL(2,\mathbb{R})$ Lie group}

We start by writing down the WZW  model on the $GL(2,\mathbb{R})$ group.
The Lie algebra $ gl(2, \mathbb{R}) = sl(2, \mathbb{R}) \oplus u(1)$ is generated by the set $(T_{1}, T_{2}, T_{3}, T_{4})$ with the following commutation
rules:
\begin{eqnarray}\label{3.1}
[T_{1} , T_{2}]~=~2T_{2},~~~~~[T_{1} ,T_{3}]~=~-2T_{3},~~~~~[T_{2} , T_{3}]~=~T_{1},~~~~~~~[T_{4},.]~=~ 0,
\end{eqnarray}
where $T_{4}$ is a central generator. As mentioned in the above, one can obtain the (undeformed) standard WZW model
from \eqref{2.5} by considering $\kappa=1$ and setting
$\eta = \tilde{A}=0 $ in formula \eqref{2.8}.
To make WZW model one needs a bilinear form  $\Omega_{ab}$
so that it satisfies the following condition \cite{nappi1993wess}:
\begin{eqnarray}\label{2}
f_{ab}^{\;\;d} \;\Omega_{dc}+ f_{ac}^{\;\;d} \;\Omega_{db}\;=\;0.
\end{eqnarray}
A bilinear form on the  $gl(2,\mathbb{R})$ Lie algebra defined by commutation relations \eqref{3.1} can be obtained by \eqref{2}, giving \cite{Exact}
{\small \begin{eqnarray}\label{3}
	\Omega_{ab}~=~\left( \begin{tabular}{cccc}
	$2\lambda$ & $0$ & $0$ & $0$\\
	$0$ & $0$ & $\lambda$ & $0$ \\
	$0$ & $\lambda$ & $0$ & $0$ \\
	$0$ & $0$ & $0$ & $\rho$\\
	\end{tabular} \right),
	\end{eqnarray}}
where $\lambda$, $\rho$ are some real constants.

In order to calculate the left-invariant one-forms we parameterize the $GL(2,\mathbb{R})$ group manifold
with the coordinates $x^{\mu} =(x, y, u, v)$, therefore the elements of the  $GL(2,\mathbb{R})$ can be written as
\begin{eqnarray}\label{3.5}
g\;=\; e^{y T_{2}} ~ e^{x T_{1}} ~ e^{u T_{3}}~ e^{v T_{4} },
\end{eqnarray}
then, using \eqref{3.1}, \eqref{3.5} together with \eqref{2.6} one can obtain the components of left-invariant one-forms, giving us\cite{Exact}
\begin{eqnarray}\label{A.10}
L^{\hspace{-0.5mm}1}_{\pm}&=&u e^{-2 x} \partial_{_{\pm}} y +\partial_{_{\pm}} x,\nonumber\\
L^{\hspace{-0.5mm}2}_{\pm}&=&e^{-2x} \partial_{_{\pm}} y,\nonumber\\
L^{\hspace{-0.5mm}3}_{\pm}&=& - u^2 e^{-2x} \partial_{_{\pm}}y -2 u
\partial_{_{\pm}} x +  \partial_{_{\pm}}u,\nonumber\\
L^{\hspace{-0.5mm}4}_{\pm}&=& \partial_{_{\pm}} v.
\end{eqnarray}
Thus, using these, the WZW  model on the $Gl(2, \mathbb{R})$ is worked out \cite{Exact}:
\begin{eqnarray}
S_{_{WZW}}(g) &=& \frac{1}{2} \int d \sigma^+ d \sigma^-~\Big[\rho~ \partial_{_{+}} v \partial_{_{-}} v
+2\lambda\big(\partial_{_{+}} x \partial_{_{-}} x+ e^{-2x} \partial_{_{+}}
u \partial_{_{-}} y\big)\Big].~~~\label{A.11}
\end{eqnarray}
By comparing the above action with the original $\sigma$-model
of the form
\begin{eqnarray}\label{3.7}
S=\frac{1}{2}\int_{_{\Sigma}}\!d\sigma^+  d\sigma^- \big(G_{_{\mu\nu}} +B_{_{\mu\nu}}\big)\partial_{+}x{^{^\mu}}
\partial_{-}x^{^{\nu}},
\end{eqnarray}
one concludes that the background metric $G_{_{\mu\nu}}$ and the antisymmetric $B$-field have the following forms:
\begin{eqnarray}
ds^2 &=& \rho d v^{2} + 2d x^{2}~ + 2 e^{-2x}~ dy~ du,\label{3.8}\\
B&=&  e^{-2x}~ du \wedge dy.\label{3.8.1}
\end{eqnarray}
Here we have assumed that $\lambda =1$. In the following, we will also use this assumption.

\subsection{\label{subSec.III.2} Classical r-matrices for the $gl(2,\mathbb{R})$ Lie algebra}
Classical r-matrices for $gl(2,\mathbb{R})$ Lie algebra were, firstly, found in \cite{Multiparameter}.
There, the Lie bialgebras structures and corresponding classical r-matrices were classified in two multiparametric inequivalent classes.
In this work, by using automorphism group of the $gl(2,\mathbb{R})$ Lie algebra and we will classify all inequivalent classical r-matrices as the solutions of (m)CYBE. We show that that classical r-matrices are split into twelve inequivalent classes.

Given Lie algebra $\G$ with the basis $\{T_a\}$  we define an element $r\in \G \otimes \G$ in the following form
\begin{eqnarray}\label{3.9}
r=r^{ab}~T_a \otimes T_b,
\end{eqnarray}
where $r^{ab}$ is an antisymmetric matrix with real entries.
The linear operator $\mathbf{R}$ associated to a classical r-matrix has an important role in the deformation process of the WZW model.
Accordingly, one may define \cite{Heisenberg}
\begin{eqnarray}\label{3.10}
\mathbf{R}(T_{c})=<r~, ~1\otimes T_{c}> = r^{ab} \Omega_{bc} ~ T_{a}.
\end{eqnarray}
Considering the relation
\begin{eqnarray}\label{3.12}
\mathbf{R}(T_{a}) = {\mathbf{R}_a}^{b} ~ T_{b},
\end{eqnarray}
and then comparing \eqref{3.10} and \eqref{3.12} one obtains that
\begin{eqnarray}\label{3.13}
{\mathbf{R}_a}^{b} =  r^{bc} \Omega_{ca}.
\end{eqnarray}
Using \eqref{3.12} and \eqref{3.13} together with \eqref{2}, one can rewrite the (m)CYBE \eqref{2.3} in the following form \cite{Heisenberg}
\begin{eqnarray}\label{3.14}
{f_{de}}^c r^{da} r^{eb}+{f_{de}}^a r^{db} r^{ec}+{f_{de}}^b r^{dc} r^{ea}
-\omega {f_{de}}^c \Omega^{da} \Omega^{eb}=0.
\end{eqnarray}
For simplicity, one may use the representations
$({\cal X}_{a})_{b}^{~c}=-{f_{ab}}^c$ and $(\mathcal{Y}^{c})_{ab}=- {f_{ab}}^c$ to obtain a matrix form of the above formula, giving \cite{Heisenberg}
\begin{eqnarray}\label{3.18}
r \mathcal{Y}^{b} r+ r (\mathcal{X}_{a} r^{ab})- (r^{ba} \mathcal{X}_{a}^{t}) r=-\omega(\Omega^{-1} \mathcal{Y}^{b} \Omega^{-1}).
\end{eqnarray}
Here, the superscript ``t'' means transposition of the matrix.
In order to calculate the r-matrices for a given Lie algebra $\G$ we need to solve equation \eqref{3.18}.
But to determine inequivalent r-matrices we should use the automorphism group of Lie algebra $\G$.
The automorphism transformation $A$ on the basis $\{T_a\}$ of $\G$ is given by $T'_a = {A}(T_a) =  {A}_a^{~b}~ T_b$,
where $\{T'_a\}$ are the changed basis by the automorphism ${A}$ obeying the same commutation relations as $\{T_a\}$.
As proved in \cite{Heisenberg} for a Lie algebra $\G$ with transformation $A \in Aut (\G)$,
two r-matrices $r$ and $r^\prime$ as solutions of the (m)CYBE are said to be equivalent
if the following transformation holds\footnote{Note that the YB deformed WZW model \eqref{2.5} is invariant under the automorphism
transformation \cite{Heisenberg}.} \cite{Heisenberg}
\begin{eqnarray}\label{3.19}
r = A^t~ r^\prime~A.
\end{eqnarray}
This equation helps us to classify the inequivalent r-matrices for the $gl(2,\mathbb{R})$.
Before proceeding to do this let us find the general automorphism of the $gl(2,\mathbb{R})$ which preserves the commutation rules \eqref{3.1}.
The automorphism transformation of $gl(2,\mathbb{R})$ which preserves the commutation rules \eqref{3.1} is given by \cite{{christodoulakis2002automorphisms},{rezaei2010complex}}
\begin{eqnarray}
T'_1 &=& (1-\frac{a}{c} \sqrt{bc}) T_1-\frac{1}{c}(ab + 2\sqrt{bc}) T_2+a T_3,\nonumber\\
T'_2 &=& (\frac{a}{2c} +\frac{a^2}{4c^2} \sqrt{bc}) T_1-(\frac{a^2 b}{4c^2}-\frac{1}{c} -\frac{a}{c^2} \sqrt{bc}) T_2+\frac{a^2}{4 c} T_3,\nonumber\\
T'_3 &=& -\sqrt{bc} T_1+bT_2-c T_3,\nonumber\\
T'_4 &=& d T_4,\label{3.20}
\end{eqnarray}
where $a, b, c, d$ are arbitrary real numbers.

For solving the (m)CYBE \eqref{3.18} for $gl(2,\mathbb{R})$ Lie algebra we consider  $r^{ab}$ as following form:
\begin{eqnarray}\label{3.25}
r^{ab}=\left(\begin{tabular}{cccc}
0 & $m_1$ & $m_2$ & $m_3$ \\
-$m_1$ & 0  & $m_4$ & $m_5$ \\
-$m_2$ & -$m_4$ & 0 & $m_6$ \\
-$m_3$ & -$m_5$ & -$m_6$ & 0 \\
\end{tabular}\right),
\end{eqnarray}
where  $m_1, \cdots ,m_6$ are real constants.
Now puttig  \eqref{3.25} into  relation \eqref{3.18} and then using \eqref{3.1} and \eqref{3},
the general  form solution of \eqref{3.18} split into six r-matrices. The solutions contain the constants $\omega$ and $m_1, \cdots ,m_6$ and are given as follows:
{\scriptsize \begin{eqnarray*}
		r_{_1}=\left(\begin{tabular}{cccc}
			0 &  $\frac{{m_4}^2}{4m_2}$ & $m_2$ & $\frac{-m_4m_6}{2m_2}$ \\
			$-\frac{{m_4}^2}{4m_2}$ & 0  &$m_4$  &  $\frac{-{m_4}^2m_6}{4{m_2}^2}$ \\
			-$m_2$ & -$m_4$ & 0 & $m_6$ \\
			$\frac{m_4m_6}{2m_2}$ &  $\frac{{m_4}^2m_6}{4{m_2}^2}$& -$m_6$ & 0 \\
		\end{tabular}\right), r_{_2}=\left(\begin{tabular}{cccc}
		0 & $m_1$  & 0 & $m_3$ \\
		-$m_1$ & 0  & $\Delta_{2}$ & $\frac{2m_1m_3}{\Delta_{2}}$ \\
		0 & -$\Delta_{2}$  & 0 & 0 \\
		-$m_3$  &- $\frac{2m_1m_3}{\Delta_{2}}$ &0 & 0 \\
	\end{tabular}\right),  r_{_3}=\left(\begin{tabular}{cccc}
	0 & $m_1$  & $m_2$ &0 \\
	- $m_1$ & 0  & $\Delta_{3}$ & 0 \\
	-$m_2$ & -$\Delta_{3}$  & 0 & 0 \\
	0 &0&0 & 0 \\
\end{tabular}\right),
\end{eqnarray*}}
{\scriptsize \begin{eqnarray}\label{3.26}
		r_{_4}=\left(\begin{tabular}{cccc}
		0 &  $\frac{-m_5\Delta_{1}}{m_6}$ & $\Delta_{1}$ & $m_3$ \\
		$\frac{m_5\Delta_{1}}{m_6}$ & 0  & $\frac{-2m_3\Delta_{1}}{m_6}$ & $m_5$ \\
		$-\Delta_{1}$ & $\frac{2m_3\Delta_{1}}{m_6}$ & 0 &$m_6$ \\
		$-m_3$ & -$m_5$ & -$m_6$ & 0 \\
	\end{tabular}\right), ~r_{_5}=\left(\begin{tabular}{cccc}
	0 & 0  & 0 & $m_3$ \\
	0 & 0  & 0 & $m_5$ \\
	0 & 0  & 0 & $m_6$ \\
	-$m_3$ &-$m_5$ &-$m_6$ & 0 \\
\end{tabular}\right),~r_{_6}=\left(\begin{tabular}{cccc}
0 & $m_1$  & 0 &0 \\
-$m_1$ & 0  & 0 & $m_5$ \\
0 &0  & 0 & 0 \\
0 &-$m_5$ &0 & 0 \\
\end{tabular}\right),~~
	\end{eqnarray}}
where $\Delta_{1} =\frac{1}{2} \sqrt{\frac{-\omega m_6^{2} }{m_5 m_6+m_3^{2}}}$,
$\Delta_{2} ={ \sqrt{-\omega }} $  and $\Delta_{3} = {\sqrt{-\omega  +4m_1 m_2}} $.\\\\
\begin{center}
	\small {{{\bf Table 1.}~  The YB deformed backgrounds of the  $GL(2, \mathbb{R})$ WZW model}}
	{\scriptsize
		\renewcommand{\arraystretch}{1.4}{
			\begin{tabular}{|p{2cm}|l|l|} \hline \hline
				{\small Background symbol }	 & {\small Backgrounds including metric and $B$-field} & {\small Comments}\\ \hline
				
				& $ds^{2}=2 dx^{2}- \eta^{2}(2+\rho) e^{-4x} dy^{2}+ 2 e^{-2x} dy du+\rho dv^{2}$ & \\
				{$GL(2,\mathbb{R})_{_i}^{(\eta,\tilde{A},\kappa)}$} & $B=\kappa e^{-2x}~ du \wedge dy+\tilde{A}\rho  e^{-2x} dv \wedge dy  $ & $\omega=0$\\&&\\\hline
				
				& $ds^{2}=2 dx^{2}-2 \eta^{2} e^{-4x} dy^{2}+ 2 e^{-2x} dy du+\rho dv^{2} $&  \\
				{$GL(2,\mathbb{R})_{_{ii}}^{(\eta,\kappa)}$}& $B=\kappa e^{-2x}~ du \wedge dy $& $\omega=0$\\& &  \\\hline
				
				& $ds^{2}=\frac{1}{1+2\rho \eta^{2}}\Big[2 dx^{2}+\rho dv^{2}-4\rho \eta^{2}u^{2} e^{-4x} dy^{2}-8\rho \eta^{2}u e^{-2x} dx dy\Big]+ 2 e^{-2x} dy du$ &  \\ $GL(2,\mathbb{R})_{_{iii}}^{(\eta,\tilde{A},\kappa)}$& $B=\kappa e^{-2x}~ du \wedge dy+\frac{2\rho\tilde{A}}{1+2\rho \eta^{2}}  ue^{-2x}dv \wedge dy $ & $\omega=0$\\& &  \\\hline
				
				& $ds^{2}=\rho dv^{2} +2\Big(1-2 \eta^{2}(2+\rho )u^{2}\Big) dx^{2}+ 2 e^{-2x}\Big(1+ \eta^{2} u^{2}(2+\rho)\Big) dy du $ & \\
				$GL(2,\mathbb{R})_{_{iv}}^{(\eta,\tilde{A},\kappa)}$&$-\eta^{2} (\rho+2)\big[du^{2} + u^{4} e^{-4x} dy^{2}+4 u^{3} e^{-2x} dx dy-4 u dx du \big]$
				&\\
				& $B= e^{-2x}(\kappa+2\tilde{A}u  )~ du \wedge dy+\tilde{A}\Big[2 u^{2}e^{-2x}dy \wedge dx +\rho u^{2}e^{-2x} dy \wedge dv+2\rho u dx \wedge dv\Big]$&  $\omega=0$ \\&&  \\\hline
				
				& $ds^{2}=\rho dv^{2} +2(1-2 \eta^{2}\rho u^{2}) dx^{2}- \rho \eta^{2}\Big[ du^{2}+ u^{4}e^{-4x}dy^{2}+4  u^{3} e^{-2x}  dx dy-4  u dx du\Big] $&\\
				$GL(2,\mathbb{R})_{_{v}}^{(\eta,\tilde{A},\kappa)}$&$+ 2 e^{-2x} (1+ \rho \eta^{2}  u^{2} )dy du $ & $\omega=0$\\
				& $B=\kappa e^{-2x}~ du \wedge dy +\rho \tilde{A}\Big[ u^{2}e^{-2x} dy \wedge dv+2u dx \wedge dv \Big] $& \\&& \\\hline
				
				& $ds^{2}=\rho dv^{2}+ 2(1-4 \eta^{2} u^{2}) dx^{2}-2 \eta^{2}\Big[ du^{2}+u^{4}e^{-4x}dy^{2}- 4 u dx du+4 u^{3} e^{-2x}  dx dy\Big]$&\\
				$GL(2,\mathbb{R})_{_{vi}}^{(\eta,\tilde{A},\kappa)}$&$+ 2 e^{-2x} (1+2 \eta^{2}  u^{2} )dy du $ & $\omega=0$\\
				& $B= e^{-2x}\big(\kappa+2\tilde{A}u\big)~ du \wedge dy+2\tilde{A} u^{2}e^{-2x}dy \wedge dx $& \\&&  \\\hline

	\end{tabular}}}
\end{center}

\begin{center}
	\small {{{\bf Table 1.}~  Continued}}
	{\scriptsize
		\renewcommand{\arraystretch}{1.4}{
			\begin{tabular}{|p{2cm}|l|l|} \hline \hline
				{\small Background symbol }	 & {\small Backgrounds including metric and $B$-field} & {\small Comments}\\ \hline
				
				& $ds^{2}=2 dx^{2}- \eta^{2}\rho e^{-4x} dy^{2}+ 2 e^{-2x} dy du+\rho dv^{2}$ & \\
				{$GL(2,\mathbb{R})_{_{vii}}^{(\eta,\tilde{A},\kappa)}$} & $B=\kappa e^{-2x}~ du \wedge dy+\tilde{A}\rho  e^{-2x} dv \wedge dy  $ & $\omega=0$\\&&\\\hline
				
				& $ds^{2}=\rho (1+\omega\eta^{2}) dv^{2}+ 2(1+\omega\eta^{2}) dx^{2}+2\omega \eta^{2}u^{2}e^{-4x}dy^{2} + 2 e^{-2x} dy du+4\omega \eta^{2}ue^{-2x} dx dy$& \\
				$GL(2,\mathbb{R})_{_{viii}}^{(\eta,\tilde{A},\kappa)}$& $B= e^{-2x}\big(\kappa -\tilde{A}\sqrt{-\omega}\big)~ du \wedge dy+2\tilde{A}\sqrt{-\omega}u e^{-2x}~ dx \wedge dy$&  $\omega <0$\\& & \\\hline
				
				& $ds^{2}=\frac{1-\eta^{2}}{1-4\eta^{2}}\Big\{2(1-\frac{2\eta^{2} (2+ \rho)}{1+2\rho\eta^{2}}u^{2}) dx^{2}+\frac{2e^{-2x}}{1+2 \rho \eta^{2}}\Big[\eta^{2}u^{2}(2+\rho)+1-\eta^{2}(2-\rho)\Big] dy du\Big\} $&\\& $-\frac{(1-\eta^{2}) \eta^{2}(2+ \rho  )}{(1-4\eta^{2})(1+2 \rho \eta^{2})}\Big[e^{-4x} (1-u^{2})^{2} dy^{2}+4ue^{-2x}(u^{2}-1)dx dy-4u dx du +du^{2}\Big]+\frac{\rho(1-\eta^{2})}{1+2 \rho \eta^{2}}dv^{2}$ & \\$GL(2,\mathbb{R})_{_{ix}}^{(\eta,\tilde{A},\kappa)}$& & $\omega=-1$\\
				& $B=( \kappa-\frac{2\tilde{A}(1-\eta^{2})}{1-4\eta^{2}}) ue^{-2x}~ du \wedge dy-\frac{2\tilde{A} (1-\eta^{2})}{1-4\eta^{2}} u^{2} e^{-2x} dy \wedge dx $& \\
				& $+\frac{\tilde{A}(1-\eta^{2}) \rho}{1+2 \rho\eta^{2}}\Big[2udx \wedge dv + e^{-2x}(-u^{2}+1) dv \wedge dy\Big]$&\\&&  \\\hline
				
				& $ds^{2}= \frac{1+\omega \eta^{2}}{1-\eta^{2}(4\omega +2 \rho )}\Big\{2\Big(1-2\rho \eta^{2}-2 \eta^{2}u^{2}(2\omega+ \rho )\Big)dx^{2}+e^{-4x}\Big[2 \eta^{2}(2\omega-\rho)u^{2}$&\\
				& $- \eta^{2}(2\omega+ \rho )(1+u^{4})\Big]dy^{2}+4 u e^{-2x}\Big[\eta^{2}(2\omega- \rho)-\eta^{2}(2\omega+ \rho)u^{2}\Big]dx dy$&\\
				& $+2e^{-2x}\Big[1- \eta^{2}(2\omega+ \rho)+ \eta^{2}u^{2}(2\omega+ \rho )\Big]dy du+8\eta^{2}\sqrt{\omega}\rho ue^{-2x}dy dv+8\eta^{2}\sqrt{\omega}\rho dx dv$&\\
				$GL(2,\mathbb{R})_{_{x}}^{(\eta,\tilde{A},\kappa)}$& $+ \rho(1-4\omega\eta^{2})dv^{2}-\eta^{2}(2\omega+\rho)du^{2}+4\eta^{2}u(2\omega+\rho)dx du\Big\}$ &\\& &$\omega >0$\\
				& $B=\Big[ \kappa-\frac{2\sqrt{\omega}\tilde{A}(1+\omega\eta^{2})}{1-\eta^{2}(4 \omega +2 \rho)}u\Big] e^{-2x}~ du \wedge dy-\frac{2\sqrt{\omega}\tilde{A} (1+\omega\eta^{2})}{1-\eta^{2}(4\omega +2 \rho)}u^{2} e^{-2x}dy \wedge dx$& \\& $+\frac{ \rho \tilde{A}(1+\omega\eta^{2})}{1-\eta^{2}(4\omega +2 \rho)}\Big[e^{-2x}(1+u^{2})dy \wedge dv+2u dx \wedge dv\Big]$&\\& & \\\hline

				& $ds^{2}=\frac{1+\eta^{2}}{1+4\eta^{2}}\Big\{2\Big(1-\frac{2\eta^{2}(2+ \rho)}{1-2 \rho\eta^{2}}u^{2}\Big) dx^{2}+\frac{2e^{-2x}}{1-2 \rho \eta^{2}}\Big[-\eta^{2}u^{2}(2+\rho)+\Big(1+\eta^{2}(2+\rho)\Big)\Big] dy du\Big\}$ &\\
				& $-\frac{(1+\eta^{2}) \eta^{2}(2+ \rho  )}{(1+4\eta^{2})(1-2 \rho \eta^{2})}\Big[e^{-4x} (1+u^{2})^{2} dy^{2}+4ue^{-2x}(u^{2}+1)dx dy-4u dx du +du^{2}\Big]+\frac{\rho(1+\eta^{2})}{1-2 \rho \eta^{2}}dv^{2}$ &  \\ $GL(2,\mathbb{R})_{_{xi}}^{(\eta,\tilde{A},\kappa)}$& & $\omega=1$\\
				& $B=\Big[ \kappa+\frac{2\tilde{A}(1+\eta^{2})}{1+4\eta^{2}}u\Big] e^{-2x}~ du \wedge dy+\frac{2\tilde{A} (1+\eta^{2})}{1+4\eta^{2}} u^{2} e^{-2x} dy \wedge dx $ &\\& $+\frac{\tilde{A}(1+\eta^{2}) \rho}{1-2 \rho\eta^{2}}\Big[-2udx \wedge dv+ e^{-2x}(u^{2}+1) dv \wedge dy\Big]$&\\& & \\\hline
				
				& $ds^{2}= \frac{1+\omega \eta^{2}}{1+\eta^{2}(-4\omega +2 \rho )}\Big\{2\Big(1+2 \rho \eta^{2}+2 \eta^{2}u^{2}(2\omega- \rho )\Big)dx^{2}+e^{-4x}\Big[2 \eta^{2}(2\omega+ \rho)u^{2}+ \eta^{2}(2\omega- \rho )$ &\\& $\times(1+u^{4})\Big]dy^{2}+4 u e^{-2x}\Big[\eta^{2}(2\omega+ \rho)+\eta^{2}(2\omega- \rho)u^{2}\Big]dx dy$ &\\& $+2e^{-2x}\Big[1-\eta^{2}(2\omega-\rho)- \eta^{2}u^{2}(2\omega- \rho )\Big]dy du-8\eta^{2}\sqrt{-\omega}\rho ue^{-2x}dy dv-8\eta^{2}\sqrt{-\omega}\rho dx dv$&\\
				$GL(2,\mathbb{R})_{_{xii}}^{(\eta,\tilde{A},\kappa)}$& $+ \rho(1-4\omega\eta^{2})dv^{2}+\eta^{2}(2\omega-\rho)du^{2}-4\eta^{2}u(2\omega-\rho)dx du\Big\}$ &$\omega <0$\\& & \\
				& $B=\Big[ \kappa+\frac{2 \tilde{A}\sqrt{-\omega}(1+\omega\eta^{2})}{1-\eta^{2}(4\omega-2 \rho)}u\Big] e^{-2x}~ du \wedge dy+\frac{\tilde{A}(1+\omega\eta^{2})}{1-\eta^{2}(4\omega-2 \rho)}\Big[2 \rho u dv \wedge dx+ \rho  e^{-2x}(u^{2}-1)dv \wedge dy$ &\\& $+2\sqrt{-\omega}e^{-2x}u^{2}dy \wedge dx\Big] $&\\&&\\\hline

	\end{tabular}}}
\end{center}

In the following, using the automorphisms transformation \eqref{3.20}
and also \eqref{3.19}, we determine representations of all inequivalent  r-matrices
of \eqref{3.26}. Indeed, they are split into twelve inequivalent classes as follows:\\\\
{\bf Theorem 3.1.}~{\it Any r-matrix of the $gl(2, \mathbb{R})$ Lie algebra as a solution of the (m)CYBE belongs
just to one of the following twelve inequivalent classes}\footnote{
In Ref. \cite{Borsato1},  in order to classify the YB deformations of the $AdS_3 \times S^3$ string, the CYBE has been solved for the Lie algebra
${sl(2,\mathbb{R})}_L\oplus {su(2)}_L\oplus {sl(2,\mathbb{R})}_R\oplus {su(2)}_R$.
There, authors have considered the basis $\{S_0, S_+, S_-\}$ for the $sl(2,\mathbb{R})$ with the commutation relations	
$[S_0 , S_{\pm}] = \pm S_{\pm},~[S_+, S_-] = 2 S_0$,  and $T_a,~ (a = 1, 2, 3)$ for the ${su(2)}$ with $[{T_a} , {T_b}]=-\epsilon_{{abc}}T_c$.
In both cases of ${sl(2,\mathbb{R})}$ and ${su(2)}$, they have used a bar to distinguish the right copy of the algebra
from left copy. In their calculations, they have focused on the subalgebra generated by the generators $\{S_0, S_+, {\bar S}_0, {\bar S}_-, T_1, {\bar T}_2\}$, and
have ignored the transformations generated by  $\{S_-, {\bar S}_+\}$.
For the algebra ${sl(2)}_L\oplus {sl(2)}_R\oplus {su(2)}_L\oplus {su(2)}_R$ with the generators  $\{S_0, S_+, {\bar S}_0, {\bar S}_-, T_1, {\bar T}_2\}$,
it has been obtained ten non-Abelian $\bf R$-matrices as the solutions of the CYBE with $\omega=0$.
We know that $gl(2,\mathbb{R}) = sl(2,\mathbb{R})\oplus u(1)$ is embedded inside $sl(2,\mathbb{R})\oplus su(2)$, therefore,
with dimensional reduction from six to four, we expect that one can obtain the $r$-matrices of Theorem 3.1 (only cases $\omega=0$) from those of \cite{Borsato1}.
By checking this we found out that the $\bf R$-matrices of those
will be only in agreement with the $r$-matrices $r_{_{i}}, r_{_{ii}}, r_{_{iii}},  r_{_{iv}}, r_{_{vi}}$ and $r_{_{vii}}$ of ours.
Thus, the $r$-matrix $r_{_{v}}$ of ours cannot be concluded by reducing the $\bf R$-matrices of those.
In fact, we have one more solution from those of \cite{Borsato1}.
}
\begin{eqnarray*}
	{r_{_i}}&=& T_{_1} \wedge T_{_3} + T_{_3} \wedge T_{_4},~~~~~~~~~~~~~~~~~~~\nonumber\\
	{r_{_{ii}}}&=& T_{_1}  \wedge T_{_3},\nonumber\\~~~~~~~~~~~~~~~~~~~
	{r_{_{iii}}}&=& T_{_1}  \wedge T_{_4},\nonumber\\~~~~~~~~~~~~~~~~~~~
	{r_{_{iv}}}&=& T_{_1}  \wedge T_{_2} + T_{_2}  \wedge T_{_4},\nonumber\\~~~~~~~~~~~~~~~~~~~
	{r_{_{v}}}&=& T_{_2}  \wedge T_{_4},\nonumber\\
	{r_{_{vi}}}&=&  T_{_1}  \wedge T_{_2},\nonumber\\
	{r_{_{vii}}}&=&  T_{_3} \wedge T_{_4},~~~~~~~~~~~~~~~~~~~\nonumber\\
	{r_{_{viii}}}&=& \sqrt{-\omega} ~T_{_2} \wedge T_{_3},\nonumber\\
	{r_{_{ix}}}&=& -T_{_1} \wedge T_{_2} +  T_{_1}  \wedge T_{_3}+T_{_2} \wedge T_{_4}+T_{_3} \wedge T_{_4},\nonumber\\
	{r_{_{x}}}&=& \sqrt{\omega} (-T_{_1} \wedge T_{_2} +T_{_1} \wedge T_{_3}) +T_{_2} \wedge T_{_4} - T_{_3}  \wedge T_{_4},\nonumber\\
	{r_{_{xi}}}&=&T_{_1} \wedge T_{_2} +T_{_1} \wedge T_{_3} -T_{_2} \wedge T_{_4} + T_{_3}  \wedge T_{_4},\nonumber\\
	{r_{_{xii}}}&=& \sqrt{-\omega} (T_{_1} \wedge T_{_2} +T_{_1} \wedge T_{_3})-T_{_2} \wedge T_{_4} - T_{_3}  \wedge T_{_4}.\nonumber
\end{eqnarray*}
Now using Eq. \eqref{3}, \eqref{3.12} and \eqref{3.13} one can find all linear $\mathbf{R}$-operators related to the inequivalent r-matrices and then
calculate the deformed currents $J_{\pm}$ and also the YB deformed WZW models.

\subsection{Backgrounds for YB deformations of the $GL(2, \mathbb{R})$ WZW model}

In this subsection we find all linear $\mathbf{R}$-operators corresponding to the inequivalent $r$-matrices  of Theorem 3.1. Then
we obtain the deformed currents $J_{\pm}$ from Eq. \eqref{2.8}. After,
using \eqref{2.5} we obtain all YB deformed backgrounds of the $GL(2,\mathbb{R})$  WZW model.
It is reminded that the symbol of each background, e.g. $GL(2,\mathbb{R})_{_i}^{(\eta,\tilde{A},\kappa)}$, indicates the YB
deformed background derived by  r-matrix $r_i$ ; roman numbers $i$, $ii$ etc. distinguish between several possible
deformed backgrounds of the WZW model, and the parameters $(\eta,\tilde{A},\kappa)$ indicate the deformation
ones of each background.
Notice that all deformed backgrounds include three parameters $(\omega,\tilde{A},\eta)$ except for $GL(2,\mathbb{R})_{_{ii}}^{(\eta,\kappa)}$.
The deformed backgrounds including metric and $B$-field are summarized in Table 1.\\

Here contrary of $H_{4}$ case \cite{Heisenberg},
none of the backgrounds of YB deformed $GL(2, \mathbb{R})$ WZW models can be related to the $GL(2, \mathbb{R})$ WZW model
(equations \eqref{3.8} and \eqref{3.8.1}), because the Killing symmetries
of the deformed metrics are different from those of equation \eqref{3.8}.
Before closing this section, let us discuss the conformal invariance conditions of the deformed models.
In Ref. \cite{Borsato2} it has been identified that a necessary and sufficient condition for the $\eta$-model to have a standard supergravity
background as target space is that an algebraic condition on the $r$-matrix.  There, it has been referred to as the unimodularity condition
that is
\begin{eqnarray}\label{4.4.1}
r^{ab}~ [T_a , T_b] =0.
\end{eqnarray}
In the procedure of YB deformation, the initial input for construction of the deformed backgrounds is the r-matrix.
When a r-matrix satisfies the unimodularity condition \eqref{4.4.1},
the YB deformed background is a solution to standard supergravity. If not, the background becomes a solution to the generalized supergravity equations.
Let us now look at the unimodularity condition on the solutions of (m)CYBE for the $gl(2, \mathbb{R})$.
Using the condition \eqref{4.4.1} together with \eqref{3.1} we find that only the $r$-matrices ${r_{_{iii}}}, {r_{_{v}}}$
and ${r_{_{vii}}}$ of Theorem 3.1 are unimodular, while the rest denote non-unimodular $r$-matrices.
So, we expect that the deformed backgrounds by the aforementioned unimodular $r$-matrices can
be satisfied the standard supergravity equations (the one-loop beta function equations \cite{callan1985strings}).
By looking at the conformal invariance conditions, we find that the backgrounds generated by the $r$-matrices ${r_{_{iii}}}, {r_{_{v}}}$
and ${r_{_{vii}}}$  satisfy the one-loop beta function equations if the deformation parameters $\eta, {\tilde A}$ vanish and $\kappa=1$.
The same condition happens for the rest of backgrounds generated by non-unimodular matrices.

\section{YB deformed models as original ones of non-Abelian T-dual $\sigma$-models }

In this section we shall show that all YB deformed WZW models of Table 1 can be obtained from a
Poisson-Lie T-dual $\sigma$-model constructed on a $2+2$-dimensional manifold ${\M}$ with the two-dimensional non-Abelian Lie group
acting freely on ${\M}$. As we will see, the dual Lie group is considered to be
Abelian.
Before we proceed to investigate this case further, let us briefly review the construction of
Poisson-Lie T-dual $\sigma$-models in the presence of spectator fields \cite{Klim1,Klim2,Sfetsos1}.
Since the Poisson-Lie duality is based on the concepts of the Drinfeld double, it is necessary to define the Drinfeld double group $D$.
A Drinfeld double \cite{Drinfeld} is simply a Lie group $D$ whose Lie algebra $\D$
admits a decomposition $\D =\G \oplus {\tilde \G}$ into a pair of sub-algebras maximally isotropic
with respect to a symmetric ad-invariant non-degenerate bilinear form $<.~,~.>$.
The dimension of sub-algebras have to be equal. We furthermore consider $G$ and $\tilde G$ as a pair of maximally isotropic subgroups corresponding
to the subalgebras $\G$ and $\tilde \G$,
and choose a basis in each of the sub-algebras as
$T_{{_a}} \in \G$  and ${\tilde T}^{{^a}} \in {\tilde \G}, a = 1, ..., dim~G$, such that
\begin{eqnarray}\label{5.1}
<T_{{_a}} ,  T_{{_b}}> = <{\tilde T}^{{^a}} ,  {\tilde T}^{{^b}}> =0,~~~~~~~~<T_{{_a}} , {\tilde T}^{{^b}}>  = {\delta}_{_{a}}^{{~b}}.
\end{eqnarray}
The basis of the two sub-algebras
satisfy the commutation relations
\begin{eqnarray}\label{5.2}
[T_a , T_b] = {f^{c}}_{ab} ~T_c,~~~~~[{\tilde T}^{a} , {\tilde T}^{b}] ={{\tilde f}^{ab}}_{~~c} ~{\tilde T}^{c},~~~~
[T_a , {\tilde T}^{b}] = {{\tilde f}^{bc}}_{~~a} {T}_c + {f^{b}}_{ca} ~{\tilde T}^{c},
\end{eqnarray}
where ${f^{c}}_{ab}$ and $\tilde f^{ab}_{~~c}$ are structure constants of $\G$ and $\tilde \G$, respectively.
Noted that the Lie algebra structure defined by relation \eqref{5.2} is called Drinfeld double $\D$.

Consider now a non-linear $\sigma$-model for the $d$ field variables
$X^{^{M}} = (x^{\mu} , y^\alpha)$, where $x^\mu$'s,  $\mu = 1, . . . , dim~G$ represent the coordinates of Lie group $G$ acting freely on
the manifold $\M \approx O \times G$, and $y^\alpha,~\alpha = 1, \cdots , d-dim~G$ are
the coordinates of the orbit $O$ of $ G$ in  ${\M}$.
A remarkable point is that the coordinates $y^\alpha$ do not participate in the Poisson-Lie  T-duality transformations
and are therefore called spectator fields \cite{Sfetsos1}.
The corresponding $\sigma$-model action has the form
\begin{eqnarray}\label{5.3}
S = \frac{1}{2} \int d\sigma^{+}  d\sigma^{-} \hspace{-6mm}&&\Big[E_{_{ab}}(g,y^{\alpha})~
R_{+}^a \;R_{-}^b + \phi^{{(1)}}_{a \beta}(g,y^{\alpha})  R_{+}^a \partial_{-} y^{\beta}+
\phi^{{(2)}}_{\alpha b}(g,y^{\alpha}) \partial_{+} y^{\alpha} R_{-}^b\nonumber\\
~~&&+\phi_{_{\alpha\beta}}(g,y^{\alpha})
\partial_{+} y^{\alpha} \partial_{-} y^{\beta} \Big].
\end{eqnarray}
where $R_{\pm}^a$ are the components of the right-invariant Maurer-Cartan one-forms which are constructed by means of
an element $g$ of the Lie group ${G}$ as
\begin{eqnarray}\label{5.4}
R_{\pm}  = (\partial_{\pm} g g^{-1})^a ~ T_a = R_{\pm}^a~ T_a=  \partial_{\pm} x^{\mu}~ R_{\mu}^{~a} ~ T_a.
\end{eqnarray}
As shown, the couplings ${{E}_{_{ab}}}, \phi^{{(1)}}_{a \alpha}, \phi^{{(2)}}_{\alpha b} $ and $\phi_{_{\alpha\beta}}$ may depend on all variables $x^\mu$ and $y^\alpha$.

Similarly we introduce another $\sigma$-model for the $d$ field variables ${\tilde X}^{^{M}} =({\tilde x}^{\mu} , y^\alpha)$, where
${\tilde x}^{\mu}$'s parameterize an element ${\tilde g}\in {\tilde G}$, whose dimension is, however,
equal to that of $G$, and the rest of the variables are the same $y^\alpha$'s used in \eqref{5.3}.
We consider the components of the right-invariant Maurer-Cartan forms on
${\tilde G}$ as $( \partial_{\pm} \tilde g \tilde g^{-1})_a={\tilde R}_{{\pm}_a}=\partial_{\pm} {\tilde x}^{\mu} {\tilde R}_{\mu a}$.
In this case, the corresponding action takes the following form
\begin{eqnarray}\label{5.5}
\tilde S = \frac{1}{2} \int d\sigma^{+}  d\sigma^{-}\hspace{-6mm}&&\Big[{{{\tilde E}}^{{ab}}}(\tilde g,y^{\alpha})~
{\tilde R}_{+_{a}}{\tilde R}_{-_{b}}+{\tilde \phi}^{\hspace{0mm}{(1)^{ a}}}_{~~~\beta}(\tilde g,y^{\alpha}) ~{\tilde R}_{+_{a}}\partial_{-} y^{\beta}+
{\tilde \phi}^{\hspace{0mm}{(2)^{ b}}}_{\alpha}(\tilde g,y^{\alpha}) \partial_{+} y^{\alpha} ~{\tilde R}_{-_{b}}\nonumber\\
&&+{\tilde \phi}_{_{\alpha\beta}}(\tilde g,y^{\alpha}) \partial_{+} y^{\alpha} \partial_{-} y^{\beta}\Big].
\end{eqnarray}
The $\sigma$-models \eqref{5.3} and \eqref{5.5} will be dual to each other in the sense of Poisson-Lie
T-duality \cite{{Klim1},{Klim2}} if the associated Lie algebras $\G$ and ${\tilde \G}$ form a the Lie algebra $\D$.
There remains to relate the couplings ${{E}_{_{ab}}}, \phi^{{(1)}}_{a \beta}, \phi^{{(2)}}_{\alpha b} $
and $\phi_{_{\alpha \beta}}$ in \eqref{5.3} to ${{{\tilde E}}^{{ab}}}, {\tilde \phi}^{\hspace{0mm}{(1)^{ a}}}_{~~~\beta},
{\tilde \phi}^{\hspace{0mm}{(2)^{ b}}}_{\alpha}$ and ${\tilde \phi}_{_{\alpha \beta}}$ in \eqref{5.5}.
It has been shown that \cite{Klim1,Klim2,Sfetsos1} the various couplings in the
$\sigma$-model action \eqref{5.3} are restricted to be
\begin{eqnarray}\label{5.6}
{{E}} &=& \big(E^{{-1}}_{0}+ \Pi\big)^{-1},~~~~~~~~~~~
\phi^{{(1)}} = {{E}}~E^{{-1}}_{0}~F^{^{(1)}},\nonumber\\
\phi^{{(2)}}&=& F^{^{(2)}}~ E^{{-1}}_{0}~{{E}},~~~~~~~~~~~~~~~~
\phi = F -F^{^{(2)}}~\Pi~{{E}}~E^{{-1}}_{0}~F^{^{(1)}},
\end{eqnarray}
where the new couplings $E_{0}, F^{^{(1)}}, F^{^{(2)}}$ and $F$ may be at most functions of
the variables $y^{\alpha}$ only. In equation \eqref{5.6},
$\Pi(g)$ defined by $\Pi^{^{ab}}(g) = b^{^{ac}}(g)~ (a^{-1})_{_{c}}^{^{~b}}(g)$ is the Poisson structure on $G$ so that matrices
$a(g)$ and $b(g)$ are defined as follows:
\begin{eqnarray}\label{5.7}
g^{-1} T_{{_a}}~ g &=& a_{_{a}}^{^{~b}}(g) ~ T_{{_b}},\nonumber\\
g^{-1} {\tilde T}^{{^a}} g &=& b^{^{ab}}(g)~ T_{{_b}}+(a^{-1})_{_{b}}^{^{~a}}(g)~{\tilde T}^{{^b}}.
\end{eqnarray}
Eventually, the relationship between the couplings of the dual action and the original one is given by \cite{Klim1,Klim2,Sfetsos1}
\begin{eqnarray}\label{5.8}
{{\tilde E}} &=& \big(E_{0}+ {\tilde \Pi}\big)^{-1},~~~~~~~~~~
{\tilde \phi}^{{(1)}} =  {{\tilde E}}~F^{^{(1)}},\nonumber\\
{\tilde \phi}^{{(2)}} &=& - F^{^{(2)}} ~{{\tilde E}},~~~~~~~~~~~~~~~~~
{\tilde \phi}           = F-F^{^{(2)}} ~{{\tilde E}}~F^{^{(1)}}.
\end{eqnarray}
Analogously, one can define matrices ${\tilde a} (\tilde g), {\tilde b} (\tilde g)$ and ${\tilde \Pi} (\tilde g)$ by just replacing the untilded symbols
by tilded ones.
As we will see, the Poisson-Lie T-duality approach in the presence of spectators helps us to
construct the non-Abelian T-dual spaces of the YB deformations of the $Gl(2, \mathbb{R})$ WZW models of Table 1.
It's worth mentioning that in Ref. \cite{Exact}, the $Gl(2, \mathbb{R})$ WZW model has been derived from a dual pair of
$\sigma$-models related by Poisson-Lie symmetry, in such a way that the WZW model as original model has been constructed on
a $2+2$-dimensional manifold ${\M} \approx O \times G$, where $G=A_2$ as a two-dimensional real non-Abelian Lie group
acts freely on ${\M}$.
Below as an example, the non-Abelian T-dualization of the YB deformed background $GL(2,\mathbb{R})_{_i}^{(\eta,\tilde{A},\kappa)}$ is discussed in detail by using the formulation mentioned above.

\subsection{Non-Abelian T-dual space of the YB deformed background $GL(2,\mathbb{R})_{_i}^{(\eta,\tilde{A},\kappa)}$}

\subsubsection{The original model}

The original model is constructed on $2+2$-dimensional manifold ${\M} \approx O \times G$ in which ${G}$ is considered to be
the Lie group $A_2$ whose Lie algebra is denoted by ${\cal A}_2$, while $O$ is the orbit of $G$ in ${\M}$.
We use the coordinates $\{x, y\}$ for the $A_2$, and employ $y^\alpha =\{u, v\}$ for the orbit $O$. In what follows we shall show
the background of original model is equivalent to the YB deformed background $GL(2,\mathbb{R})_{_i}^{(\eta,\tilde{A},\kappa)}$.
As mentioned earlier,  having  Drinfeld doubles one can construct the Poisson-Lie T-dual $\sigma$-models on them.
The Lie algebra of the semi-Abelian double $({\cal A}_2 , 2{\cal A}_1)$ is defined by the following non-zero Lie brackets
\begin{eqnarray}\label{5.9}
[T_1 , T_2]~=~2T_2,~~~~~[T_1 ~, ~{\tilde T}^2]=-2{\tilde T}^2,~~~~~[T_2 ~, ~{\tilde T}^2]=2{\tilde T}^1.
\end{eqnarray}
where $\{T_1 , T_2\}$ and $\{{\tilde T}^1 , {\tilde T}^2\}$ are the basis of ${\cal A}_2$ and $2{\cal A}_1$, respectively.
In order to calculate the components of right invariant one-forms
$R_{\pm}^a$ on the  $A_2$ we parameterize an element of $A_2$ as
\begin{eqnarray}\label{5.10}
g~=~e^{-x T_1}~e^{y T_2}.
\end{eqnarray}
Then, $R_{\pm}^a$'s are derived to be of the form
\begin{eqnarray}\label{5.11}
R^{1}_{\pm}=-\partial_{\pm} x,~~~~~~~~~R^{2}_{\pm}=e^{-2x} \partial_{\pm} y.
\end{eqnarray}
To achieve a $\sigma$-model with the background $GL(2,\mathbb{R})_{_i}^{(\eta,\tilde{A},\kappa)}$
one has to choose the spectator-dependent matrices in the following form
\begin{eqnarray}
		E_{0_{ab}}&=&\left( \begin{tabular}{cc}
			$2$ & 0  \\
			0 &$ - \eta^{2}(\rho+2)$  \\
			
		\end{tabular} \right),~~~~~~~~~~F^{(1)}_{a \beta }=\left( \begin{tabular}{cc}
			0 & 0  \\
			$ (1-\kappa)$ &$ - \tilde A \rho$ \\
		\end{tabular} \right),\nonumber\\
		F^{(2)}_{ \alpha b }&=&\left( \begin{tabular}{cc}
			0 & $ (1+\kappa)$   \\
			0 &$ \tilde A \rho$  \\
			
		\end{tabular} \right),~~~~~~~~~~~~~~~F_{\alpha \beta }=\left( \begin{tabular}{cc}
			0 & 0  \\
			0 &$  \rho$  \\
		\end{tabular} \right).\label{5.12}
\end{eqnarray}
Since the dual Lie group, $2A_1$, has assumed to be Abelian, it follows from the second relation of \eqref{5.7} that $b^{ab}(g)=0$; consequently,
$\Pi^{ab}(g)=0$. Using these and employing \eqref{5.6} one can construct the action \eqref{5.3} on the manifold ${\M} \approx O \times G$.
The corresponding background including metric and antisymmetric two-form field are given by
\begin{eqnarray}
ds^{2}&=&2 dx^{2}- \eta^{2}(\rho+2) e^{-4x} dy^{2}+ 2 e^{-2x} dy du+\rho dv^{2},\nonumber\\
B&=&\kappa e^{-2x}~ du \wedge dy+\tilde{A}\rho  e^{-2x} dv \wedge dy, \label{5.13}
\end{eqnarray}
which is nothing but the YB deformed background $GL(2,\mathbb{R})_{_i}^{(\eta,\tilde{A},\kappa)}$ as was represented in Table 1.
Thus, we showed that the background  $GL(2,\mathbb{R})_{_i}^{(\eta,\tilde{A},\kappa)}$
can be considered as original model from a dual pair of $\sigma$-models related by Poisson-Lie symmetry.
In this manner one can obtain the spectator-dependent matrices for all backgrounds of Table 1.
The results for the spectator-dependent matrices are summarized in Table 2.

\subsubsection{The dual model}

The dual model is constructed on a $2+2$-dimensional manifold
$\tilde {\M} \approx O \times \tilde {G}$
with two-dimensional Abelian Lie group ${\tilde {G}}=2A_1$ acting freely on it.
In the same way to construct out the dual $\sigma$-model we parameterize the corresponding Lie
group (Abelian Lie group $2A_1$) with  coordinates ${\tilde x}^\mu = \{{\tilde x} , {\tilde y}\}$.
In order to calculate the components of the right invariant one-forms
on the dual Lie group we parametrize an element of the group as
\begin{eqnarray}\label{5.14}
\tilde g=e^{\tilde x {\tilde T}^{1}}e^{\tilde y {\tilde T}^{2}}.
\end{eqnarray}
We then obtain
\begin{eqnarray}\label{5.15}
\tilde R_{\pm 1}=\partial_{\pm}\tilde x,~~~~~~~~~ \tilde R_{\pm 2}=\partial_{\pm} \tilde y.
\end{eqnarray}
Utilizing relation \eqref{5.7} for untilded quantities we get
{\small \begin{eqnarray}\label{5.16}
		\tilde \Pi_{ab}=\left( \begin{tabular}{cc}
			0 & $-2\tilde y $  \\
			$2\tilde y $ &0 \\
			
		\end{tabular} \right).
\end{eqnarray}}
Now  inserting \eqref{5.12} and \eqref{5.16} into equations \eqref{5.8}
one can obtain dual couplings, giving us
{\small \begin{eqnarray*}
		\tilde	E^{ab}=\frac{1}{{\Delta}}\left( \begin{tabular}{cc}
			$\frac{ \eta^{2}(2+\rho)}{2}$ & ${-\tilde y}$  \\
			${\tilde y}$ &$-1 $  \\
			
		\end{tabular} \right),~~~~~~~~~~~	{\tilde \phi}^{\hspace{0mm}{(1)^{ a}}}_{~~~\beta}=\frac{1}{{\Delta}}\left( \begin{tabular}{cc}
			${\tilde y 	(\kappa-1)}$ & ${ \tilde A \tilde y\rho}$  \\
			${(\kappa-1)}$ &${ \tilde A \rho}$  \\
		\end{tabular} \right),
\end{eqnarray*}}
{\small \begin{eqnarray}
		{\tilde \phi}^{\hspace{0mm}{(2)^{ b}}}_{\alpha}=\frac{1}{{\Delta}}\left( \begin{tabular}{cc}
			${-\tilde y 	 (1+\kappa)}$ & ${ (1+\kappa)}$  \\
			$-{ \tilde A \tilde y\rho}	$ &${ \tilde A \rho}$  \\
			
		\end{tabular} \right),~~~~~~~~~~~{\tilde \phi}_{_{\alpha \beta }}=\frac{1}{{\Delta}}\left( \begin{tabular}{cc}
			${ (1-\kappa^{2})}$& $-{ \tilde A (1+\kappa) \rho}$ \\
			${\tilde A (1-\kappa) \rho}$ &$  \rho {\Delta}- { \tilde A^{2} \rho^{2}} $  \\
		\end{tabular} \right),
\end{eqnarray}}
where ${\Delta}=\eta^{2}(\rho + 2)-2 \tilde y^{2} $.
Finally, inserting the above results into action \eqref{5.5}, the corresponding metric and the antisymmetric tensor field are worked out to be
\begin{eqnarray}
d\tilde s^{2}&=&\frac{1}{\Delta} \Big[\frac{1}{2} \eta^{2}(\rho +2) d{\tilde x}^{2}-d {\tilde y}^{2}+(1-\kappa^{2}) du^{2}+
(\rho {\Delta} - {\tilde A}^{2} \rho^{2})dv^{2}- 2 {\tilde y} d{\tilde x}du \nonumber\\
&&~~~~~+2  {\kappa} d {\tilde y} du+ 2\tilde A \rho (d\tilde y~ dv  - \kappa dv du)\Big],\nonumber\\
\tilde B&=& \frac{1}{\Delta} \Big[{\tilde y}~ d \tilde y \wedge d\tilde x + {\tilde y  \kappa } ~d\tilde x \wedge du +  du \wedge d \tilde y
+ {\tilde y  \tilde A \rho }~ d\tilde x \wedge dv +{\tilde A  \rho}~ dv \wedge du\Big].
\end{eqnarray}


\begin{landscape}
\begin{center}
	\small {{{\bf Table 2.}~  Spectator-dependent background matrices}}
	{\scriptsize
		\renewcommand{\arraystretch}{1.5}{
			\begin{tabular}{|p{2cm}|c|c|c|c|l|} \hline \hline
				
				{\small Background symbol}	&
				$E_{0}$     &
				$ F^{(1)}$  &
				$ F^{(2)}$   &
				$ F$         &
				\small Comments\\
				\hline
				
				\multirow{ 3}{*}{$GL(2,\mathbb{R})_{_i}^{(\eta,\tilde{A},\kappa)}$} &
				\multirow{ 3}{*}{\footnotesize $\left(\begin{array}{ll} 2& 0\\0 & - \eta^{2}(2+\rho)\\ \end{array} \right)$}&
				\multirow{ 3}{*}{\footnotesize$\left(\begin{array}{ll}  0&0\\ 1-\kappa&- \tilde A \rho \\\end{array} \right)$}&
				\multirow{3}{*}{\footnotesize$\left(\begin{array}{ll}   0& 1+\kappa\\ 0& \tilde A \rho\\\end{array} \right)$}&
				\multirow{3}{*}{\footnotesize$\left(\begin{array}{ll}   0&0\\0&\rho \\\end{array} \right)$}&\\
					&&&&&\\
					&&&&&\\
					\hline
				
				\multirow{ 3}{*}{$GL(2,\mathbb{R})_{_{ii}}^{(\eta,\kappa)}$} &
				\multirow{ 3}{*}  	{\footnotesize $\left(\begin{array}{ll} 2 & 0         \\0 & -2\eta^{2}  \\  \end{array} \right)$}&
				\multirow{ 3}{*}    {\footnotesize$\left(\begin{array}{ll}  0 & 0          \\ 1-\kappa& 0 \\ \end{array} \right)$}&
				\multirow{ 3}{*}    {\footnotesize$\left(\begin{array}{ll}  0 & 1+\kappa \\ 0&0\\             \end{array} \right)$}&
				\multirow{ 3}{*}    {\footnotesize$\left(\begin{array}{ll}  0 &0            \\0&\rho          \\\end{array} \right)$}& \\
						&&&&&\\
						&&&&&\\
						\hline
				
\multirow{ 3}{*}  {$GL(2,\mathbb{R})_{_{iii}}^{(\eta,\tilde{A},\kappa)}$}~~~ &  	
\multirow{ 3}{*}  {\footnotesize $\left(\begin{array}{l l} \frac{2}{1+2\rho \eta^{2}}& \frac{4 \rho\eta^{2}u}{1+2\rho \eta^{2}}\\\frac{4 \rho\eta^{2}u}{1+2\rho \eta^{2}} & \frac{-4 \rho\eta^{2}u^{2}}{1+2\rho \eta^{2}}\\ \end{array} \right)$}&
\multirow{ 3}{*} {\footnotesize$\left(\begin{array}{l l}0&0\\ 1-\kappa&\frac{-2 \rho \tilde A u}{1+2\rho \eta^{2}}
\\\end{array} \right)$}&
\multirow{ 3}{*}{\footnotesize$\left(\begin{array}{ll} 0& 1+\kappa\\ 0&\frac{2 \rho \tilde A u}{1+2\rho \eta^{2}}\\\end{array} \right)$}&
\multirow{ 3}{*}{\footnotesize$\left(\begin{array}{ll} 0&0          \\ 0&\frac{\rho}{1+2\rho \eta^{2}}             \\\end{array} \right)$}&\\
                        &&&&&\\
						&&&&&\\
						\hline
				
&  & & & &
{\footnotesize $P=\eta^{2}(\rho+2)$} \\
&
\multirow{ 3}{*}{\footnotesize $\left(\begin{array}{ll} 2 (1-2 u^2 P) & P_{+} \\ P_{-} & u^4 P\\ \end{array} \right)$}&
\multirow{ 3}{*}{\footnotesize$\left(\begin{array}{ll} -2 u P & -2 \tilde A \rho u        \\ P_1-P_2&\rho u^{2} \tilde A \\\end{array} \right)$}&
\multirow{ 3}{*}{\footnotesize$\left(\begin{array}{ll}  -2 u P& P_1+P_2       \\  2 \tilde A \rho u&-\rho u^{2} \tilde A\\\end{array} \right)$}&
\multirow{ 3}{*}{\footnotesize $\left(\begin{array}{ll} -P &0\\0&\rho \\\end{array} \right)$}&
\\
{$GL(2,\mathbb{R})_{_{iv}}^{(\eta,\tilde{A},\kappa)}$} &  & & & &
{\footnotesize $P_{\pm}=2u^{2}(u P \pm\tilde A)$} \\

&  & & & &
{\footnotesize $P_1=1+2 u^2 P$} \\

&  & & & &
{\footnotesize$P_2=\kappa+2\tilde A  u$}\\\hline

&\multirow{ 3}{*}{\footnotesize $\left(\begin{array}{ll} 2 \Sigma_{_-} & 2u^{3}\rho \eta^{2}\\2u^{3} \eta^{2}\rho & -\eta^{2}\rho u^{4}\\ \end{array} \right)$}&
\multirow{ 3}{*}{\footnotesize$\left(\begin{array}{ll} -2 \eta^{2} \rho u&-2 \tilde A \rho u\\ \Sigma_{_+}-\kappa  &\rho u^{2}  \tilde A\\\end{array} \right)$}&
\multirow{ 3}{*}{\footnotesize$\left(\begin{array}{ll} -2 \eta^{2} \rho u& \Sigma_{_+} +\kappa \\ 2\tilde A \rho u&-\rho u^{2} \tilde A\\\end{array} \right)$}&
\multirow{ 3}{*}{\footnotesize$\left(\begin{array}{ll} -\eta^{2 } \rho&0\\0&\rho\\\end{array} \right)$}&\\

{$GL(2,\mathbb{R})_{_{v}}^{(\eta,\tilde{A},\kappa)}$} &  & & & &
{\footnotesize $\Sigma_{+}=1+ \rho \eta^{2} u^{2}$} \\

&  & & & &
{\footnotesize $\Sigma_{-}=1-2 \rho \eta^{2} u^{2}$} \\ \hline
			
&  & & & &
{\footnotesize $\Lambda_{_\pm}=2\eta^{2} u \pm \tilde A$} \\
&
\multirow{ 3}{*}{\footnotesize $\left(\begin{array}{l l} 2 P_4 & 2 u^2 \Lambda_{_+}\\ 2 u^2 \Lambda_{_-}  & -2\eta^{2} u^{4}\\ \end{array} \right)$}&
\multirow{ 3}{*}{\footnotesize$\left(\begin{array}{l l}-4 \eta^{2}u&0\\ P_3-P_2 &0\\\end{array} \right)$}&
\multirow{ 3}{*}{\footnotesize$\left(\begin{array}{ll}-4 \eta^{2}u& P_3+P_2\\ 0&0\\\end{array} \right)$}&
\multirow{ 3}{*}{\footnotesize$\left(\begin{array}{ll}-2\eta^{2 }&0\\0&\rho
						\\\end{array} \right)$}&
\\
{$GL(2,\mathbb{R})_{_{vi}}^{(\eta,\tilde{A},\kappa)}$} &  & & & &
{\footnotesize $P_{3}=1+2 \eta^{2} u^{2}$} \\

&  & & & &
{\footnotesize $P_4=1-4 \eta^{2} u^{2}$} \\\hline

\multirow{ 3}{*}{$GL(2,\mathbb{R})_{_{vii}}^{(\eta,\tilde{A},\kappa)}$} &
\multirow{ 3}{*}{\footnotesize $\left(\begin{array}{ll} 2& 0\\0 & - \eta^{2}\rho\\ \end{array} \right)$}&
\multirow{ 3}{*}{\footnotesize$\left(\begin{array}{ll}  0&0\\ 1-\kappa&- \tilde A \rho \\\end{array} \right)$}&
\multirow{3}{*}{\footnotesize$\left(\begin{array}{ll}   0& 1+\kappa\\ 0& \tilde A \rho\\\end{array} \right)$}&
\multirow{3}{*}{\footnotesize$\left(\begin{array}{ll}   0&0\\0&\rho \\\end{array} \right)$}&\\
&&&&&\\
&&&&&\\
\hline

	&\multirow{ 3}{*}{\footnotesize $\left(\begin{array}{l l} 2 (1+\omega \eta^{2}) & \Sigma_{1}-\Sigma_{2}\\ \Sigma_{1} +\Sigma_{2} & u \Sigma_{1} \\ \end{array} \right)$}&
	\multirow{ 3}{*}{\footnotesize$\left(\begin{array}{l l}0&0\\ 1 -\Sigma_{3} &0
		\\\end{array} \right)$}&
	\multirow{ 3}{*}{\footnotesize$\left(\begin{array}{ll}0& \Sigma_{3}+1\\ 0&0\\\end{array} \right)$}&
	\multirow{ 3}{*}{\footnotesize$\left(\begin{array}{ll}0&0\\0&\rho(1+\omega \eta^{2} )
		\\\end{array} \right)$}&\\
	
	{$GL(2,\mathbb{R})_{_{viii}}^{(\eta,\tilde{A},\kappa)}$} &  & & & &
	{\footnotesize $\Sigma_{1} = -2\omega \eta^{2} u$} \\
	
	&  & & & &
	{\footnotesize $\Sigma_{2}=2 \tilde A\sqrt{-\omega} u$ }\\	&  & & & &
	{\footnotesize$\Sigma_{3}=\kappa- \tilde A\sqrt{-\omega}$}\\\hline

	\end{tabular}}}
\end{center}
\end{landscape}
\newpage
\begin{landscape}
\begin{center}
	
	\scriptsize {{{\bf Table 2.}~   Continued}}
	{\scriptsize
		\renewcommand{\arraystretch}{1}{
			\begin{tabular}{|p{1.7cm}|l|l|l|l|l|} \hline \hline
				
				{\tiny Background symbol }	&
				$~~~~~~~ E_{0}$     &
				$ ~~~~~~~ F^{(1)}$  &
				$~~~~~~~ F^{(2)}$   &
				$~~~~~~~ F$         &
				\tiny Comments
				\\ \hline
				&  & & & &
				{\tiny $ \Lambda_{_{1_{-}}}=\frac{(1-\eta^{2}) }{(1-4  \eta^{2})} $} \\
				 &  & & & &
				 {\tiny  $\gamma_{_{1_{-}}} = \frac{\eta^{2}(2 +  \rho )}{1+2 \rho\eta^{2}} $ } \\
				
				&  & & & &
				{\tiny $\varGamma_{_{2_{-}}}=2 u^{2}\tilde A \Lambda_{_{1_{-}}}$} \\
				
				&  & & & &
				{\tiny$\varGamma_{_{3_{-}}}=\frac{\rho \tilde A (1-\eta^{2})}{1+2 \rho \eta^{2}} $}\\
				&
				\multirow{ 3}{*}	{\tiny $\left(\begin{array}{l l} {\tiny 2(1-2u^{2}\gamma_{_{1_{-}}})\Lambda_{_{1_{-}}} } & {\tiny \varGamma_{_{1_{-}}}-\varGamma_{_{2_{-}}}}\\\ {\tiny \varGamma_{_{1_{-}}}+\varGamma_{_{2_{-}}}} &  \delta_{_-}^{2}\delta_{_{1_{-}}}\\ \end{array} \right)$}&
				\multirow{ 3}{*}{\tiny$\left(\begin{array}{l l}{\tiny 2u \delta_{_{1_{-}}} } &-2u\varGamma_{_{3_{-}}}\\  \gamma_{_{2_{-}}} -\varGamma_{_-} &-\delta_{_-}\varGamma_{_{3_{-}}}
					\\\end{array} \right)$}&
				\multirow{ 3}{*}{\tiny$\left(\begin{array}{ll}2u \delta_{_{1_{-}}}& \varGamma_{_-}+ \gamma_{_{2_{-}}}\\ 2u\varGamma_{_{3_{-}}} &\delta_{_-}\varGamma_{_{3_{-}}}\\\end{array} \right)$}&
				\multirow{ 3}{*}{\tiny $\left(\begin{array}{ll}\delta_{_{1_{-}}} &0\\0&\frac{\varGamma_{_{3_{-}}}}{\tilde A}
					\\\end{array} \right)$}&
				\\
				{{\tiny $GL(2,\mathbb{R})_{_{ix}}^{(\eta,\tilde{A},\kappa)}$ }}
					&  & & & &
				{\tiny$  \varGamma_{_-}=\kappa-2 u \tilde A \Lambda_{_{1_{-}}}$}\\
					&  & & & &
				{\tiny $\delta_{_-}=1-u^{2}$}\\
					&  & & & &
				{\tiny$  \varGamma_{_{1_{-}}}= -2u  \delta_{_-}\gamma_{_{1_{-}}}\Lambda_{_{1_{-}}} $}\\
					&  & & & &
				{\tiny $\delta_{_{1_{-}}} = - \gamma_{_{1_{-}}} \Lambda_{_{1_{-}}} $}\\	
                     &  & & & &
				{\tiny $\gamma_{_{2_{-}}} = -\delta_{_{1_{-}}} [u^{2}+$}\\
					&  & & & &
				{\tiny $ \frac{1-\eta^{2}(2-\rho) }{\eta^{2}(2 +  \rho )}]$}\\
					
				 \hline
			
				&  & & & &
{\tiny $\lambda_{_\pm}=2 \omega\pm\rho$}\\
				&  & &  & &
			{\tiny $\Lambda_{_{2_{+}}}=\frac{1+\omega\eta^{2}}{1-2\eta^{2} \lambda_{_+} }$} \\
				&  & & & &
			{\tiny $\gamma_{_{3_{+}}}=2u \eta^{2}[\lambda_{_-}-\lambda_{_+} u^{2}]$} \\
			
			&  & & & &
			{\tiny$\gamma_{_{4_{+}}}=2\tilde A\sqrt{\omega}u^{2} $}\\
			&  & & & &
			{\tiny $\gamma_{_{5_{+}}}=2\lambda_{_-}u^{2}-\lambda_{_+}(1+u^{4})$}\\
				&  & & & &
			{\tiny $\gamma_{_{6_{+}}}=\Lambda_{_{2_{+}}} [1+\eta^{2}\lambda_{_+} $}\\
			&  & & & &
			{\tiny $\times (-1+u^{2})]$}\\
			&  & & & &
			
		{\tiny $\gamma_{_{7_{+}}} = 1-2  \rho \eta^{2}-2 \eta^{2} u^{2}\lambda_{_+} $} \\

			&
			\multirow{ 3}{*}		{\tiny $\left(\begin{array}{l l}2  \gamma_{_{7_{+}}} \Lambda_{_{2_{+}}} & -\Lambda_{_{2_{+}}} \gamma'_{_+}  \\ \Lambda_{_{2_{+}}} \gamma'_{_-}  & \Lambda_{_{2_{+}}}\eta^{2} \gamma_{_{5_{+}}} \\ \end{array} \right)$}&
		
			\multirow{ 3}{*}{\tiny $\left(\begin{array}{l l}- \Lambda_{_{2_{+}}} \varGamma_{_{8_{+}}} &-\Lambda_{_{2_{+}}} \varGamma'_{_+}\\ \gamma_{_{6_{+}}} -\varGamma _{_{4_{+}}} &\Lambda_{_{2_{+}}}\lambda'_{_+}
				\\\end{array} \right)$}&
			\multirow{ 3}{*}	{\tiny$\left(\begin{array}{ll}- \Lambda_{_{2_{+}}} \varGamma_{_{8_{+}}}& \varGamma _{_{4_{+}}}+ \gamma_{_{6_{+}}}\\\Lambda_{_{2_{+}}}\varGamma'_{_-} &\Lambda_{_{2_{+}}}\lambda'_{_-}\\\end{array} \right)$}&
			\multirow{ 3}{*}{\tiny$\left(\begin{array}{ll} \varGamma_{_{10_{+}}} &0\\0& \varGamma_{_{9_{+}}}
				\\\end{array} \right)$}&
			\\
			{{\tiny $GL(2,\mathbb{R})_{_{x}}^{(\eta,\tilde{A},\kappa)}$ }} &  & & & &
		{\tiny$\gamma'_{_\pm}=\gamma_{4+}\pm \gamma_{3+} $}	\\

			&  & & & &
			{\tiny $ \varGamma_{_{4_{+}}}=\kappa-2 u \tilde A\sqrt{\omega}\Lambda_{_{2_{+}}}$}\\
			&  & & & &
			{\tiny $ \varGamma_{_{5_{+}}}=4\eta^{2}\sqrt{\omega}\rho$}\\
			&  & & & &
			{\tiny $\varGamma_{_{6_{+}}}=2u\rho\tilde A$}\\
				&  & & & &
			{\tiny $\varGamma_{_{7_{+}}}= \rho  \tilde A(1+ u^{2})$}\\
			&  & & & &
			{\tiny $\varGamma_{_{8_{+}}}= 2\eta^{2}u \lambda_{_+}$}\\
				&  & & & &
			{\tiny $\varGamma_{_{9_{+}}}= \rho \Lambda_{_{2_{+}}} (1-4\omega\eta^{2})$}\\
				&  & & & &
			{\tiny $\varGamma_{_{10_{+}}}=- \eta^{2} \Lambda_{_{2_{+}}} \lambda_{_+}$}\\
			&  & & & &
			{\tiny $\varGamma'_{_\pm}=\varGamma_{_{6_{+}}}\pm \varGamma_{_{5_{+}}} $}\\
				&  & & & &
			
	    	{\tiny $\lambda'_{_\pm}=u\varGamma _{_{5_{+}}}\pm\varGamma_{_{7_{+}}}$}\\
				&  & &  & &
			\\\hline

	\end{tabular}}}
\end{center}
\end{landscape}

\newpage
\begin{landscape}
	\begin{center}
		
		\scriptsize {{{\bf Table 2.}~   Continued}}
		{\scriptsize
			\renewcommand{\arraystretch}{.5}{
				\begin{tabular}{|p{1.6cm}|l|l|l|l|l|} \hline \hline
					
					{\tiny Background symbol }	&
					$~~~~~~~ E_{0}$     &
					$ ~~~~~~~ F^{(1)}$  &
					$~~~~~~~ F^{(2)}$   &
					$~~~~~~~ F$         &
					\tiny Comments
					\\ \hline
				
					&  & & & &
					{\footnotesize{\tiny $ \Lambda_{_{1_{+}}}=-\frac{1+\eta^{2}}{1+4 \eta^{2}}  $ } } \\
					&  & & & &
					{\footnotesize {\tiny $ \delta_{_+}=1+u^{2}$ }} \\
					
					&  & & & &
					{\footnotesize {\tiny $ \gamma_{_+} =\frac{\eta^{2}(\rho +2)}{1-2 \rho\eta^{2}} $ } } \\
					
					&  & & & &
					{\footnotesize {\tiny $\varGamma_{_{3_{+}}}=\frac{ \rho \tilde A (1+\eta^{2})}{1-2 \rho \eta^{2}}   $ }}\\
						
					&
					\multirow{ 3}{*}	{\tiny $\left(\begin{array}{l l}-2(1-2u^{2}\gamma_{_+})\Lambda_{_{1_{+}}}  & \varGamma_{_{1_{+}}}+\varGamma_{_{2_{+}}}\\\ \varGamma_{_{1_{+}}}-\varGamma_{_{2_{+}}} & \delta_{_{1_{+}}} \delta_{_+}^{2} \\ \end{array} \right)$}&
					\multirow{ 3}{*}{\tiny $\left(\begin{array}{l l}2u\delta_{_{1_{+}}} &2u\varGamma_{_{3_{+}}}\\ \gamma_{_{2_{+}}} -\varGamma_{_+} &-\delta_{_+} \varGamma_{_{3_{+}}}
						\\\end{array} \right)$}&
					\multirow{ 3}{*}{\tiny $\left(\begin{array}{ll}2u \delta_{_{1_{+}}}& \varGamma_{_+}+ \gamma_{_{2_{+}}}\\ -2u\varGamma_{_{3_{+}}} &\delta_{_+} \varGamma_{_{3_{+}}}\\\end{array} \right)$}&
					\multirow{ 3}{*}{\tiny$\left(\begin{array}{ll} \delta_{_{1_{+}}} &0\\0&\gamma_{_{1_{+}}}
						\\\end{array} \right)$}&
					\\
					{{\tiny $GL(2,\mathbb{R})_{_{xi}}^{(\eta,\tilde{A},\kappa)}$}}
					&  & & & &
					{\tiny  $\delta_{_{1_{+}}}=\gamma_{_+}\Lambda_{_{1_{+}}} $} \\
					&  & & & &
					{\tiny$\varGamma_{_{1_{+}}}=-2u   \delta_{_+}\delta_{_{1_{+}}} $}\\
						&  & & & &
					{\tiny $ \varGamma_{_{2_{+}}}=-2 u^{2}\tilde A \Lambda_{_{1_{+}}}  $} \\
					&  & & & &
					{\tiny$\gamma_{_{1_{+}}} =  \frac{\rho (1+\eta^{2})}{1-2 \rho \eta^{2}}$}\\
					&  & & & &
					{\tiny$\gamma_{_{2_{+}}}= -\delta_{_{1_{+}}}[1-u^{2}$}\\
					&  & & & &
					{\tiny$+\frac{1}{\eta^{2}(2+  \rho )}]  $}\\
					&  & & & &
					{\tiny $\varGamma_{_+}=\kappa-2u \tilde A \Lambda_{_{1_{+}}}  $} \\
					\hline
					
					&  & & & &
					{\tiny $\Lambda_{_{2_{-}}}=\frac{1+\omega\eta^{2} }{1-2\eta^{2} \lambda_{_-} }$} \\

					&  & & & &
					{\tiny $\gamma_{_{3_{-}}}=2u \eta^{2}[\lambda_{_+}+u^{2}\lambda_{_-}]$} \\
					&  & & & &
					{\tiny $\gamma_{_{4_{-}}}=2\tilde A u^{2}\sqrt{-\omega} $}\\
					&  & & & &
					{\tiny $\gamma_{_{5_{-}}}=\eta^{2}\Lambda_{_{2_{-}}}  [2u^{2}\lambda_{_+} $}\\
					&  & & & &
					{\tiny$ +(1+u^{4})\lambda_{_-}]$}\\
						&  & & & &
					{\tiny $\gamma_{_{6_{-}}} =  1+2\eta^{2} (\rho+ u^{2}\lambda_{_-}) $} \\
					&  & & & &
					{\tiny $\gamma''_{_\pm}=\gamma_{_{4_{-}}}\pm \gamma_{_{3_{-}}}$}
					\\
					
					&
					\multirow{ 3}{*}	{\tiny $\left(\begin{array}{l l}2 \gamma_{_{6_{-}}}\Lambda_{_{2_{-}}} & \Lambda_{_{2_{-}}} \gamma''_{_-}  \\-\Lambda_{_{2_{-}}} \gamma''_{_+}  &\gamma_{_{5_{-}}}\\ \end{array} \right)$}&
					\multirow{ 3}{*}{\tiny$\left(\begin{array}{l l}2 u \Lambda_{_{2_{-}}} \varGamma''_{-} &\Lambda_{_{2_{-}}}\lambda''_{_-}\\  \varGamma_{_{10_{-}}} -\varGamma_{_{4_{-}}} &\Lambda_{_{2_{-}}} \Lambda'_{_-}
						\\\end{array} \right)$}&
					\multirow{ 3}{*}	{\tiny$\left(\begin{array}{ll}2u\Lambda_{_{2_{-}}} \varGamma''_{_-}&\varGamma_{_{10_{-}}} +\varGamma_{_{4_{-}}} \\ -\Lambda_{_{2_{-}}}\lambda''_{_+} &\Lambda_{_{2_{-}}} \Lambda'_{_+}\\\end{array} \right)$}&
					\multirow{ 3}{*}{\tiny$\left(\begin{array}{ll} \Lambda_{_{2_{-}}} \varGamma''_{_-} &0\\0& \Lambda_{_{2_{-}}} \varGamma_{_{9_{-}}}
						\\\end{array} \right)$}&
					\\
					{{\tiny $GL(2,\mathbb{R})_{_{xii}}^{(\eta,\tilde{A},\kappa)}$ }}
					&  & & & &
					{\tiny$\varGamma_{_{4_{-}}}=\kappa+ 2u \sqrt{-\omega} \tilde A \Lambda_{_{2_{-}}}$}\\
					&  & & & &
					{\tiny $\varGamma_{_{5_{-}}}=-2\eta^{2}\sqrt{-\omega}$}\\
					&  & & & &
					{\tiny$\varGamma_{_{7_{-}}}= 2u \rho \varGamma_{_{5_{-}}} $}\\
					&  & & & &
					{\tiny$\varGamma_{_{8_{-}}}= \rho  \tilde A( u^{2}-1)$}\\
					&  & & & &
					{\tiny$\varGamma_{_{9_{-}}}=\rho  (1-4\omega\eta^{2}) \Lambda_{_{2_{-}}}$}\\
					&  & & & &
					{\tiny$\varGamma_{_{10_{-}}}=  (1-\eta^{2}\lambda_{_-}\delta_{_+}) \Lambda_{_{2_{-}}} $}\\
						&  & & & &
					{\tiny $\Lambda' _{_\pm}=\varGamma_{_{7_{-}}}\pm \varGamma_{_{8_{-}}} $} \\
					&  & & & &
					{\tiny $$} \\
					 	&  & & & &
					 {\tiny $\varGamma'' _{_\pm}=\eta^{2} \lambda_{_\pm} $} \\
					&  & & & &
					{ $ $} \\
					&  & & & &
					{\tiny $\lambda''_{_\pm}=\varGamma_{_{6_{+}}}\pm 2\varGamma_{_{5_{-}}}$}\\
					&  & &  & &
					\\\hline

		\end{tabular}}}
	\end{center}
\end{landscape}



\section{Summary and concluding remarks}

We have obtained the inequivalent classical r-matrices for the $gl(2,\mathbb{R})$
Lie algebra as the solutions of (m)CYBE by using its corresponding automorphism transformation.
Using these we have constructed the YB deformations of the $GL(2,\mathbb{R})$ WZW model.
Our results including twelve models have been summarized in Table 1.
We have shown that each of these models can be obtained from a
Poisson-Lie T-dual $\sigma$-model in the presence of the spectator fields when the dual Lie group is considered to be
Abelian. This means that all deformed models have Poisson-Lie symmetry just as undeformed WZW model on the $GL(2,\mathbb{R})$.
In fact, Poisson-Lie symmetry has been preserved under the YB deformation.
Since all information related to the deformation of models is collected in the spectator-dependent background matrices
$E_{0}, F^{(1)}, F^{(2)}$ and $F$, it seems that will be possible
for another choice of these matrices (except ours), other integrable backgrounds can be made.
This is a question that we will address in the future
and hope to find such backgrounds.


\subsection*{Acknowledgements}

This work has been supported by the research vice chancellor of Azarbaijan Shahid Madani University under research fund No. 97/231.



\end{document}